\documentclass[pre,aps,onecolumn,superscriptadress,floatfix]{revtex4}
\usepackage{amssymb,amsmath,graphicx,bbm,color}
\allowdisplaybreaks
\usepackage{amsfonts}
\usepackage{verbatim}
\usepackage{relsize}
\usepackage{hyperref}

\usepackage{dsfont}

\usepackage{dsfont}

\newcommand{\tr}{\operatorname{tr}}

\usepackage{dsfont}

\newcommand{\beg}{\begin}

\newcommand{\trm}{\textrm}
 
\newcommand{\bgt}{\begin{itemize}}
\newcommand{\ent}{\end{itemize}}

\newcommand{\op}{\operatorname}

\newcommand{\lan}{\langle}
\newcommand{\ran}{\rangle}

\newcommand{\Tr}{\operatorname{Tr}}

\newcommand{\E}{\op{\mathbb{E}}}

\newcommand{\f}{\frac}

\newcommand{\bbm}{\begin{bmatrix}}
\newcommand{\ebm}{\end{bmatrix}}
\newcommand{\bes}{\begin{equation*}}
\newcommand{\ees}{\end{equation*}}
\newcommand{\be}{\begin{equation}}
\newcommand{\ee}{\end{equation}}
\newcommand{\beqy}{\begin{eqnarray}}
\newcommand{\eeqy}{\end{eqnarray}}
\newcommand{\beq}{\begin{eqnarray*}}
\newcommand{\eeq}{\end{eqnarray*}}
\newcommand{\one}{\mathbbm{1}}

\newcommand{\bpm}{\begin{pmatrix}}
\newcommand{\epm}{\end{pmatrix}}

\newcommand{\tred}{\textcolor{red}}
\newcommand{\tblue}{\textcolor{blue}}

\long\def\symbolfootnote[#1]#2{\begingroup
\def\thefootnote{\fnsymbol{footnote}}\footnote[#1]{#2}\endgroup}

\renewcommand\Re{\operatorname{Re}}
\renewcommand\Im{\operatorname{Im}}

\begin{document}



\title{ Statistical diagonalization of a biased random Hamiltonian: the case of the eigenvectors}

\author{Gr\'egoire Ithier}      \affiliation{Department of Physics, Royal Holloway, University of London, United Kingdom}
\author{Saeed Ascroft}      \affiliation{Department of Physics, Royal Holloway, University of London, United Kingdom}



%
%
%
%

\begin{abstract}
We present a non perturbative calculation technique providing the mixed moments of the overlaps between the eigenvectors of two large quantum Hamiltonians: $\hat{H}_0$ and $\hat{H}_0+\hat{W}$, where 
$\hat{H}_0$ is deterministic and $\hat{W}$ is random.
We apply this method to recover the second order moments or Local Density Of States in the case of an arbitrary fixed $\hat{H}_0$ and a Gaussian $\hat{W}$. 
Then we calculate the fourth order moments of the overlaps in the same setting. Such quantities are crucial for understanding the \textit{local} dynamics of a large composite quantum system. In this case, $\hat{H}_0$ is the sum of the Hamiltonians of the system subparts
 and $\hat{W}$ is an interaction term.  We test our predictions with numerical simulations.
\end{abstract}

\maketitle

\section{Introduction}
How are the eigenvalues and eigenvectors of an hermitian matrix (or Hamiltonian) $\hat{H}_0$ modified by the addition of another hermitian  matrix $\hat{W}$?
 This question is central in many areas of  science: e.g. in physics for the quantum many body problem\cite{Thouless,Mattuck,Nozieres}, 
 quantum chaos\cite{izrailev_simple_1990,Haake} and thermalisation\cite{deutsch_quantum_1991,deutschnotes,ithier_typical_2017}, the Anderson localization problem\cite{AndersonLocalization1958,lee_disordered_1985,evers_anderson_2008}, in signal processing for telecommunications  and time series analysis\cite{allez_eigenvectors_2014,bun_rotational_2015,bun_overlaps_2016}, to name a few.
Perturbation theory provides a \textit{deterministic} answer, i.e. for given $\hat{H}_0$ and $\hat{W}$, for both the spectrum and the eigenvectors, and provided the typical strength of $\hat{W}$ is much smaller than the minimum level 
spacing of $\hat{H}_0$. This approach has been the main focus in physics for a long time.
On the non perturbative side, the Bethe Ansatz\cite{Bethe1931} provides some exact diagonalization results but only for specific classes of matrices (i.e. Hamiltonians), see e.g., \cite{JMLuckQWalkers,luck2016investigation,JMLuck2017}.
If one focuses on the particular problem of finding only the spectrum of $\hat{H}_0+\hat{W}$ for \textit{arbitrary} given matrices
 $\hat{H}_0$ and $\hat{W}$, this is a very difficult problem~\cite{weyl_asymptotische_1912,klyachko,fulton,Tao2001}. However, 
 getting a \textit{probabilistic} answer for some classes of random matrices (e.g. for $W$) is possible and not less satisfactory for the physicist looking for typical properties.
For instance, Dyson's brownian motion\cite{dyson_brownianmotion_1962} provides  the spectral properties of $\hat{H}_0+\hat{W}$
but only when the matrix $\hat{W}$ is random with identically distributed Gaussian entries.
More recently, first order "free" probability theory has also provided a probabilistic 
answer regarding the global spectral properties of $\hat{H}_0+\hat{W}$ and for larger classes of matrices\cite{VoiculescuFreeConvol,VoiculescuSubordination,BianeFreeProba}, namely matrices in generic position with one another\cite{FreeProbaNote}.
Some partial results on the \textit{local} spectral properties of $\hat{H}_0 + \hat{W}$ given by the second order statistics (i.e. the correlation functions)
 have been obtained using Random Matrix Theory tools \cite{orourke_universality_2013} and the concept of second order ``freeness'' \cite{Mingo1,Mingo2,Collins3} also seems  promising to deeply understand how correlation functions combine together when summing large matrices.
 On the question of the eigenvectors, much less work has been done (see however \cite{wilkinson_brownian_1995,allez_eigenvector_2012,allez_eigenvectors_2013,allez_eigenvectors_2014,bun_rotational_2015,bun_overlaps_2016}), and the natural question is:
 what is the statistics of the overlaps (or scalar products) between eigenvectors of $\hat{H}_0$ and eigenvectors 
 of $\hat{H}_0+\hat{W}$.

In this article, we
  present a non perturbative method for calculating the mixed moments of these overlaps under generic assumptions on a deterministic $\hat{H}_0$ and a random $\hat{W}$. This method is inspired from\cite{khorunzhy_asymptotic_1996} where the authors considered moments of traces of the resolvent operator without source term (i.e. $\hat{H}_0=0$) and was also used in \cite{Capitaine2011} for investigating the behavior of eigenvalues under additive matrix deformation.
  The method we present is approximate and reminiscent of the loop equations (or Schwinger-Dyson equations) of the diagram technique in Quantum Field Theory \cite{Abrikosov,makeenko1991,makeenko1991b}. It consists in finding a self consistent approximate solution to a set of algebraic equation  verified by the mixed moments of the Green functions\cite{GreenFunctionNote}.
 Compared to other methods used for quantitatively describing the behavior of eigenvectors under matrix addition, 
 e.g. replica trick\cite{bun_rotational_2015,bun_overlaps_2016}, supersymmetric formalism\cite{mirlin_universality_1991,fyodorov_localization_1991,fyodorov_analytical_1992,mirlin_statistics_1993}, or flow equation method\cite{genway_dynamics_2013,allez_eigenvectors_2013,allez_eigenvectors_2014},
 the method we propose is technically less involved and could in principle provide access to the approximation error in a transparent way\cite{ApproxNote}. 
  The method relies on the fact that some quantities involved in the calculation are the subject of ``measure concentration'' (see, e.g., Ref.\cite{MR1387624,LedouxConcentration,chatterjee_fluctuations_2007}), in other words they are self-averaging which allows to spot easily the
 main contribution in order to derive and solve the approximate self-consistent equations verified by the Green functions mixed moments. 
 In addition, this technique does not require rotational invariance of the probability distribution of the additive term and could be used to investigate
   the effect of correlation between entries, of a non Gaussian statistics or of a band structure. Theses cases are relevant in many contexts, e.g. for the empirical covariance estimation problem\cite{allez_eigenvectors_2014} or for a two body interaction in physical systems\cite{flambaum_correlations_1996,Kota}.
As far as the second and fourth order moments of the overlaps in the Gaussian $\hat{W}$ case are concerned, the main application we will have in mind regards the dynamics of quantum systems, and in particular the process of thermalisation\cite{ithier_dynamical_2017,ithier_typical_2017}. However, our method is general and could be used to tackle more generally problems of quantum phase transitions in disordered and interacting quantum systems.

  This paper is organized as follows: 
In Sec.\ref{Section3}, we present our calculation framework and illustrate our method with the second order moment calculation by recovering the results of the GOE and GUE cases for $\hat{W}$ with an arbitrary $\hat{H}_0$. 
Then in Sec.\ref{Section4}, we focus on the fourth order statistics, i.e. the most important quantity for understanding the out of equilibrium dynamics of embedded quantum systems (as we will explain later), and
 expose the main result of this paper in the case of a Gaussian distributed $\hat{W}$ (GUE and GOE) with arbitrary $\hat{H}_0$. 
 
\section{Analytical framework for calculating the moments of the overlap coefficients}
\label{Section3}
In this section, 
 we introduce our framework and apply it to the calculation of the second order statistics of the overlaps coefficients. 
 We recall several tools useful for our calculation: the link between the overlap coefficients and the resolvent of the total Hamiltonian, the expansion of a resolvent matrix entry as a function of the interaction Hamiltonian
(in Sec.\ref{ResolventExpansion}) and a so called "decoupling" formula (in Sec.\ref{SectionDecoupling}). 
Using these tools we recover the well known result for the first order statistics of the resolvent in the case
of a Gaussian interaction (see also \cite{allez_eigenvectors_2013}). 
We infer the second order statistics of the overlaps in Sec.\ref{SecondStatOverlaps}
and compare the analytical prediction to the results of numerical simulations (Fig.\ref{Fig1}).

 \subsection{Hypotheses}
\subsubsection{Decomposing $\hat{H}$ in two parts: $\hat{H}_0$ and $\hat{W}$}
 Following the same notations as in \cite{ithier_dynamical_2017,ithier_typical_2017},
 we consider a ``bare'' Hermitian matrix $\hat{H}_0$ with eigenvectors $\{|\phi_1 \ran, ..., |\phi_N\ran \}$ and respective eigenvalues
 $\epsilon_1,...,\epsilon_N$, and a ``dressed'' matrix $\hat{H}=\hat{H}_0+\hat{W}$ (with $\hat{W}$ Hermitian)
 with eigenvectors $\{|\psi_1 \ran, ..., |\psi_N \ran \}$ and eigenvalues $\{ \lambda_1, ...,  \lambda_N \}$.
Such a separation of the total matrix $\hat{H}$ in two parts is natural (but not unique) in several contexts, e.g. when modelling
disordered quantum systems\cite{anderson_absence_1958,evers_anderson_2008} like metals or many body interacting quantum systems like
 heavy nuclei\cite{wigner_characteristic_1955,wigner_characteristics_1957} or atoms\cite{gribakina_band_1995,Gribakin1999}. 
An important focus in mesoscopic physics today is to incorporate \textit{both} disorder and interactions and try to understand their interplay, e.g. when studying the many body localization problem\cite{gornyi_interacting_2005,basko_metalinsulator_2006,huse_localization-protected_2013,schreiber_observation_2015}. 
Such separation is also relevant for modeling the effect of noise when considering the empirical estimation of a matritial quantity, e.g. some covariance between time series, with a relatively small data set\cite{allez_eigenvectors_2014,bun_rotational_2015,bun_overlaps_2016}.
 As we mainly have physical applications in mind, we will call in the following the matrices $\hat{H}_0,\hat{H}$ the bare and dressed Hamiltonians and 
 $\hat{W}$ the interaction. 

\subsubsection{Introducing randomness}
The quantities of interest regarding how the eigenvectors of $\hat{H}_0$ are modified by the addition of the extra term $\hat{W}$ are the overlap coefficients or
scalar products:  $\lan \phi_n | \psi_i \ran$. These coefficients define the transition matrix from the bare basis to the dressed one which physically speaking tells how a
bare eigenvector $|\phi_n \ran$  is \textit{hybridized} with the dressed eigenvectors. 
 Outside the perturbative limit, analytical calculation of the quantities $ \lan \phi_n | \psi_i \ran $  have proved to be difficult for \textit{deterministic} matrices.
 Now therefore if physical Hamiltonians should be considered a priori like fully deterministic matrices, it can be interesting to introduce some level of \textit{controlled} randomness in the modeling.
This was Wigner's original idea when considering the nucleus Hamiltonian like a random matrix. 
Proceeding this way, he was making a crucial step from the usual statistical physics approach where
 randomness is introduced on the state of a system\cite{MicroCanonicalNote}
towards a new statistical physics where randomness is now introduced on the \textit{nature} of the system itself.
This radical change of point of view was originally justified on heuristic grounds:
in practice, physically relevant quantities, i.e. accessible by experiments, 
do not depend much on the details of the realization of the disorder associated to the randomness. These physical observables seem to take \textit{typical} values dependent only on some conditions constraining randomness and summarizing its macroscopic properties, e.g., symmetry class: hermitian or real symmetric, spectral variance, the possibility of a block diagonal structure. 
Recently, the typicality of the quantum dynamics\cite{ithier_dynamical_2017} provided a rigorous ground for justifying such introduction of a controlled amount of randomness in the modeling of a quantum system, in this case in
the interaction Hamiltonian $\hat{W}$ between a system $S$ (Hamiltonian $\hat{H}_s$) and its environment (Hamiltonian $\hat{H}_e)$. 
This typicality property states that the reduced density matrix of $S$, i.e. the state of $S$, considered as a function $f(\hat{W})$ of a random interaction $\hat{W}$ (either Wigner band random matrix or a randomly rotated matrix), for all other parameters fixed (initial state $\varrho(0)$, densities of states of $\hat{H}_0$ and $\hat{W}$), exhibits a generalized
 central limit theorem phenomenon known  as the ``concentration of measure''\cite{LedouxConcentration,MR1387624}:
$$ f(\hat{W}) = \varrho_s(t) = \Tr_e(\varrho(t))= \Tr_e \left(U_t \varrho(0) U_t^\dagger \right)  \text{ where } U_t=e^{-iHt} \text{ and } \hat{H}=\hat{H}_s+\hat{H}_e +\hat{W},$$
is such that the variance of $f$ with respect to the probability measure on $\hat{W}$ verifies
$$\sigma_f^2 \leq \f{4 \sigma_w^2 t^2}{\hbar^2} \f{1}{\dim \mathcal{H}_e},$$
where $\sigma_w^2=\Tr(\hat{W}.\hat{W}^\dagger)/N$ is assumed to be fixed, independent of $N$.
As a consequence, when $\dim \mathcal{H}_e \to \infty$, $\sigma_f^2 \to 0$ and  the embedded system $S$ follows a typical dynamics
given by $\E[\varrho_s(t)]$\cite{ithier_dynamical_2017}. A first consequence of this property is to explain the lack of sensitivity to microscopic details of $\hat{W}$ of processes like for instance thermalisation. 
A second consequence of this property is very practical:  it allows to calculate an approximation of $\varrho_s(t)$ simply by averaging over the interaction $\hat{W}$:
 $\Tr_e (\varrho(t)) \approx  \E[\Tr_e(\varrho(t))] =\Tr_e(\E[\varrho(t)]) $. Such calculation requires the fourth moments of the overlap coefficients as we will see now.

\subsubsection{Motivation for calculating the overlap moments}
\label{Motivation}
Our main motivation in the overlap moments calculation concerns the time evolution of a quantum system coupled to a large environment and the so called ``thermalisation'' problem. It is usually argued in the litterature on this problem that the second order moments
 give complete information about the dynamics. We argue that it is actually the fourth order moments, as far as the state of the embedded system is concerned. 
Indeed, expanding the initial state on the bare eigenbasis $\varrho(0)=\sum_{m,p} c_{m,p} |\phi_m \ran \lan \phi_p | $ and the evolution operator $\hat{U}_t$ on the dressed eigenbasis:
 $\hat{U}_t =\sum_i | \psi_i \ran \lan \psi_i | e^{-i\lambda_i t}  $, one can see easily that the matrix elements of the total
density matrix $ \lan \phi_n | \varrho(t) | \phi_q \ran$ involves such quantities as 
$$\sum_{i,j} e^{-i(\lambda_i-\lambda_j)t} \lan \phi_n | \psi_i \ran   \lan \psi_i | \phi_m\ran \lan \phi_p | \psi_j\ran \lan \psi_j | \phi_q\ran.$$
After averaging and taking the large dimension limit in order to consider a continuous approximation, we are left with, on one hand, the two point density function of the dressed spectrum $p(\lambda,\lambda')$ and, on the other hand, the fourth order moments of the overlap coefficients: $ \E[ \lan \phi_n | \psi_i \ran   \lan \psi_i | \phi_m\ran \lan \phi_p | \psi_j\ran \lan \psi_j | \phi_q\ran]$. These quantities allow to calculate $\E[\varrho(t)]$ and subsequently $\varrho_s(t)\approx \Tr_e(\E[\varrho(t)])$.
This motivates our interest in the calculation of the moments of the overlaps, and in particular the fourth order ones.
%

\subsection{Calculation tools}

\subsubsection{Link between the overlaps and the Green functions}
We first remind the well know relations between the overlaps and
 the matrix elements of the resolvent operator of the dressed Hamiltonian $G_{\hat{H}}(z)=(\hat{H}-z\one)^{-1}$ in the bare eigenbasis $ \{ |\phi_1 \ran,..., |\phi_N \ran \}$:
 $G_{n,m}(z_1)=\lan \phi_n |G(z_1)|\phi_m\ran$.
 These $G_{n,m}(z)$ are similar to the familiar Green functions or propagators of quantum field theory\cite{Abrikosov} which are the matrix elements of the resolvent on the real space $|\vec{r} \ran$ basis. 
 In our case, to stay as general as possible we consider an abstract Hilbert space and the matrix elements of the resolvent are considered
 on the bare basis $\{ |\phi_1 \ran , ... , |\phi_n \ran \}$.
Using the closure relation verified by the dressed eigenbasis: $G_{n,m}(z)= \sum_{j} \lan \phi_n | \psi_j\ran \lan \psi_j |\phi_m\ran \f{1}{\lambda_j-z}$, 
we see that the overlap $ \lan \phi_n | \psi_j\ran \lan \psi_j |\phi_m\ran$ is  the residue of 
the complex function $z \mapsto G_{n,m}(z)$ at the pole $\lambda_j$.
Defining the retarded Green functions as $G_{n,m}^R(\lambda) = \lim_{\eta \to 0^+}  G_{n,m} (\lambda + i\eta) $,
 expanding the fraction with $z = \lambda + i \eta$ for $\eta \to 0^+$, 
and taking the imaginary part, we get the quantity  $\f{1}{\pi} \Im G_{n,m}^R(\lambda) =  \sum_{j} \lan \phi_n | \psi_j \ran \lan \psi_j |\phi_m\ran 
 \delta(\lambda-\lambda_j),$
which, in the case $n=m$, coincides with the Local Density Of States (LDOS) also called Strength Function in nuclear physics or spectral function in condensed matter. This function can be seen as a non perturbative extension of the so-called spectral function and can be probed experimentally, e.g. in neutron scattering experiment for the nuclear LDOS or angle resolved photoelectric emission "ARPES" for the electronic LDOS\cite{claessen_fermi-liquid_1992}.
 In order to introduce the various tools needed for calculating the moments of the overlaps, we focus first on
 the second order ones:
 $\E[\lan \phi_n | \psi_i \ran  \lan \psi_i | \phi_m \ran ]$.
%
%
 Note that in this article, 
 we will not worry about the precise shape of the probability distribution of each overlap $\lan \phi_n | \psi_i \ran$
 and if they may obey some kind of 
 generalized Porter-Thomas distribution (i.e. Gaussian distribution for the overlaps). We refer the reader to \cite{porter_fluctuations_1956,rosen_slow_1960,desjardins_slow_1960,garg_neutron_1964} for experimental evidence, \cite{deutsch_quantum_1991,deutschnotes} for some
 insight on this problem,
 \cite{allez_eigenvectors_2014} for a full derivation when $\hat{W}$ is Gaussian and
\cite{Kota} for the binary approximation which relies on a Gaussian statistics assumption for the overlaps.

Coming back to the Green functions, we will assume to be in the non perturbative regime (i.e. $\sigma_w \gg D$) so that after averaging, these Green functions
  no longer have isolated poles but a branch cut along the support of the dressed spectrum.
Each second order moment of the overlap is sampling the step height of this branch cut which can be related to the difference between the retarded Green function  $ G^R_{n,m}(\lambda) = \lim_{\eta \to 0^+} \E[G_{n,n}(\lambda+i\eta)] $ and the advanced Green function  $G^R_{n,m}(\lambda)= \lim_{\eta \to 0^+ }\E[G_{n,m}(\lambda-i \eta)]$) around an average dressed eigenvalue:
\be
\label{LinkOverlapResolvent1}
\boxed{\E[ \lan \phi_n | \psi_j \ran \lan \psi_j |\phi_m \ran ] 
\approx  \f{1}{2 i \pi} \f{1}{\rho} \lim_{\eta \to 0^+} \left( \E[ G_{n,m}(\lambda_j +i\eta)] -  
\E[ G_{n,m}(\lambda_j - i\eta)] \right) }
\ee
We thus need to calculate the averaged Green functions.
	
\subsubsection{Identifying the zero mean Green functions}
\label{LargeZexp}
This is the first step in the calculation: identifying the zero terms.
We use here the same method as in\cite{karginconcentration}  which relies on a large $|z|$ expansion of the resolvent operator: 
$$G_H(z) = - \f{1}{z} \sum_{k=0}^\infty \f{H^k}{z^k}.$$
Assuming $\hat{W}$ to be Gaussian (either GOE or GUE) and using the Wick theorem, one can show easily that $\E[H^k]$ is diagonal $\forall k \in \mathbb{N}^+$ in the bare eigenbasis (see Supp. Mat. of \cite{ithier_typical_2017}).
This implies that all extra diagonal mean Green functions are zero: $\E[G_{n,m}(z)]=0$ $\forall z$ for $n \neq m$ when $\hat{W}$ is Gaussian.
Note that finding the value of the diagonal terms is very difficult using this $1/z$ expansion and involves advanced combinatoric reasoning.
We prefer to use the following much simpler loop equation technique, which requires two sets of preliminary formulas: the expansion of the Green functions and a so-called ``decoupling'' formula.

\subsubsection{Expansion of a Green function $G_{n,m}(z)$ with respect to the interaction $\hat{W}$.}
\label{ResolventExpansion}
For the sake of completeness we remind here some well known properties.
We consider the expansion of a resolvent entry as function of the interaction $\hat{W}$. Our starting point is the identity involving the resolvent of the sum of two matrices $\hat{H}=\hat{H}_0+\hat{W}$:
$G_{\hat{H}}(z)=G_{\hat{H}_0}(z)-G_{\hat{H}_0}(z) \hat{W}G_{\hat{H}}(z)
=G_{\hat{H}_0}(z)-G_{\hat{H}}(z) \hat{W}G_{\hat{H}_0}(z)
$
which follows trivially from the resolvent definition $G_{\hat{H}}(z)(\hat{H}-z\one)=\one$ and is a propagator version of the 
Lippmann-Schwinger equation. 
This equation provides several useful well known formulas:
\begin{eqnarray}
\label{ResolventEntryExpansion}
G_{n,m}(z) & = &   \frac{1}{\epsilon_n-z} \left( \delta_{n,m}- \sum_{k =1}^N W_{n,k} G_{k,m}(z)  \right)
= \frac{1}{\epsilon_m-z} \left( \delta_{n,m}- \sum_{k =1}^N  G_{n,k}(z) W_{k,m}  \right) \\
\label{DiffResolventIdentities}
\f{\partial G_{n,p}(z)}{ \partial W_{k,l}} & = & - G_{n,k}(z) G_{l,p}(z) \\
\f{G_H(z_1)-G_H(z_2)}{z_1-z_2} & = & - G_H(z_1) G_H(z_2)  \qquad \text{which gives} \qquad \f{\partial G_{n,p}(z)}{\partial z}  =  -\lan \phi_n | G(z)^2 |\phi_p \ran 
\end{eqnarray}
where $\delta_{n,m}$ is the Kroenecker symbol. 
One should note that the expansions in Eq.\eqref{ResolventEntryExpansion} are equalities  and not approximations, i.e. they are not Taylor expansions. These equations can be considered as \textit{equations of motion}.
One can also note that a diagram perturbation calculation at order $k$ would mean to iterate the expansion process $k$ times and neglect the residual. Here, we need only a single such expansion and do \textit{not} perform any truncation. 
\subsubsection{``Decoupling'' formula.}
\label{SectionDecoupling}
This is the core tool of our method for averaging Green functions and their products: a ``decoupling'' formula,
which was previously used
 in\cite{khorunzhy_asymptotic_1996} 
 for calculating the covariance between \textit{traces} of the resolvent of random matrices without source term, i.e. $\hat{H}_0=0$. This formula consists in a cumulant expansion approach based on the following simple idea: if $\xi$ is a real random variable, and $f$ a complex value function defined on $\mathbb{R}$, then $\E[\xi f(\xi)]$ can be written as an expansion over the cumulants $\kappa_n(\xi)$ of $\xi$: 
\be
\label{DecouplingFormulaPrelim}
\E[\xi f(\xi)]= \sum_{n=0}^p   \f{\kappa_{n+1}(\xi)}{n!} \E[f^{(n)} (\xi)] +...
\ee
This formula follows easily from integration by parts.
We will consider here a generalization of this formula to the multivariate case with
$\vec{\xi}=(\xi_1,\xi_2,...,\xi_N)$ :

\begin{equation}
\label{DecouplingFormula}
\E[\xi_1 f(\vec{\xi})]= \kappa_1(\xi_1) \E[f]+\sum_j \frac{\kappa_2(\xi_1,\xi_j)}{1!} \E\left[ \frac{\partial f}{\partial \xi_j}\right]
+\sum_{j,k} \frac{\kappa_3(\xi_1,\xi_j,\xi_k)}{2!} \E\left[ \frac{\partial^2 f}{\partial \xi_j \partial \xi_k}\right] +...
\end{equation}
where the $\kappa_r(\xi_1,\xi_{i_2},...,\xi_{i_r})$ are the mixed cumulants of order $r$.
The ``decoupling'' effect is now clear: this formula  allows to relate the covariance between the input and the output of the function $f$ to the statistics of the input and the statistics of the derivatives of $f$.
 This formula simplifies when the random variables $\{\xi_1,\xi_2, ...,\xi_N\}$ form a centered Gaussian family.
 Only the second order remains, since all higher order mixed cumulants are zero: 
\begin{equation}
\label{AverageXFX}
\quad 
\boxed{ \E[\xi_1 f(\vec{\xi})]=\sum_k \kappa_2(\xi_1,\xi_k) \E\left[\f{\partial f}{\partial \xi_k}\right] \qquad \text{if the $\{\xi_1,...\xi_N\}$ are real centered Gaussian variables}.}
\end{equation}
 $\kappa_2(\xi_i,\xi_j)$ is the covariance matrix of the $\{ \xi_i \}$ family.
In this article, we will consider such a truncation of the decoupling formula in Eq.\eqref{DecouplingFormula} at order $2$, which means that we will take into account only the Gaussian behavior in the statistics of $\hat{W}$. Such simplification provides a first path for capturing 
all the phenomenon important we are interested in (in particular thermalisation\cite{ithier_typical_2017}) and is sufficient for our purpose.
Calculation with higher order cumulants or correlations between entries are more involved and will be investigated in a further publication. Note that thanks to the decoupling formula the familiar power series of perturbation theory has been changed for a cumulant series, where the terms of order higher than $2$ are \textit{exactly} zero in the Gaussian interaction case. 
%
\subsubsection{Covariance between an entry of $\hat{W}$ and a Green function}

Using the decoupling formula, we calculate the covariance between $W_{n,m}$ and $G_{p,q}(z)$, a quantity required in the next sections:
\begin{eqnarray}
\E[W_{n,m} G_{p,q}(z) ] &= & 
  \frac{ \sigma_w^2}{N}  \E\left[ \frac{\partial G_{p,q}(z)}{\partial W_{m,n}}\right]  \text{ in the GUE case,} \nonumber \\
  \label{WG}
& =&   \frac{ \sigma_w^2}{N} \left(  \E\left[ \frac{\partial G_{p,q}(z)}{\partial W_{m,n}}\right] +\E\left[ \frac{\partial G_{p,q}(z)}{\partial W_{n,m}}\right]   \right) \text{ in the GOE case,}  
\end{eqnarray}
where $\sigma_w$ is defined as the standard deviation of the spectrum of $\hat{W}$:
$\sigma_w^2 =\tr(\hat{W}^2)$ (with $\tr=\Tr/N$ the normalized trace).

\subsubsection{Loop equations for the mean Green functions}
\label{FirstStatResolvent}
We can now start the calculation of the mean of a diagonal Green function $G_{n,n}(z)$.
The method consists in the following steps:
 \begin{itemize}
\item Expand $G_{n,n}(z)$ using Eq.\eqref{ResolventEntryExpansion}:
$ (\epsilon_n -z ) G_{n,n}(z)  =   1 - \sum_{k =1}^N W_{n,k} G_{k,n}(z) $.
\item Average over the statistics of $\hat{W}$ and use the decoupling formula from Eq.\eqref{WG} in order to get
\begin{eqnarray}
 \label{GUEeq}
 (\epsilon_n-z) \E[G_{n,n}(z)] &= &
1+\f{\sigma_w^2}{N}\sum_k \E[G_{k,k}(z) G_{n,n}(z)]= 1+\sigma_w^2 \E[ m_H(z) G_{n,n}(z)]  \qquad   \text{ in the GUE case,} \\
 \label{GOEeq}
  &= &1+\sigma_w^2 \E[ m_H(z) G_{n,n}(z)]  +\f{\sigma_w^2}{N} \sum_k \E[G_{k,n}(z)^2]  \qquad  \text{ in the GOE case,}
 \end{eqnarray}
  where $m_{\hat{H}}(z)= \tr(G_{\hat{H}}(z))$ is the Stieltjes transform of $\hat{H}$. 
    \item Spot a self-averaging quantity in these last equations: $m_{\hat{H}}(z)$ has the property of being concentrated around its mean value for
$z=\epsilon+i \eta$ not too close from the support of the spectrum (i.e. $\eta \gg D$ the mean level spacing)
(see for instance Corollary 4.4.30 in \cite{agz}), since it is a sum of a large number of weakly correlated terms.
\item Neglect the fluctuations of this concentrated quantity around its mean value and identify $m_{\hat{H}}(z)$ with its mean value $\E[m_{\hat{H}}(z)]$. Derive a \textit{self-consistent} approximate equation verified by the mean Green function $\E[G_{n,n}(z_1)] $ (also called a loop or Schwinger-Dyson equation). Solving this equation in the GUE case provides the diagonal mean Green functions:
\begin{equation}
\label{GaussianResolvent}
\E[G_{n,n}(z)] 
\approx  \frac{1}{\epsilon_n-z-\sigma_w^2  m_{\hat{H}}(z) }.
\end{equation}
 \item In the GOE case, there is an extra term: 
 $\f{\sigma_w^2}{N} \sum_k \E[G_{k,n}(z)^2]$. However, we will show in Sec.\ref{ZCasesOrder2} that $\E[G_{k,n}(z)^2]=0$ for $n \neq k$,
so that the GOE result is identical to the GUE one.
\end{itemize}
It is important to notice that Eq.\eqref{GaussianResolvent} is a self-consistent equation, since the Stieltjes transform
on the r.h.s. involves diagonal Green functions for all values of $n$. One should also note that the quantity $\sigma_w^2 m_H(z)$ extends the concept of self-energy to a non perturbative situation.


\subsubsection{Second order moments of the overlaps.}
\label{SecondStatOverlaps}

\begin{figure}
\includegraphics[width=0.8\textwidth]{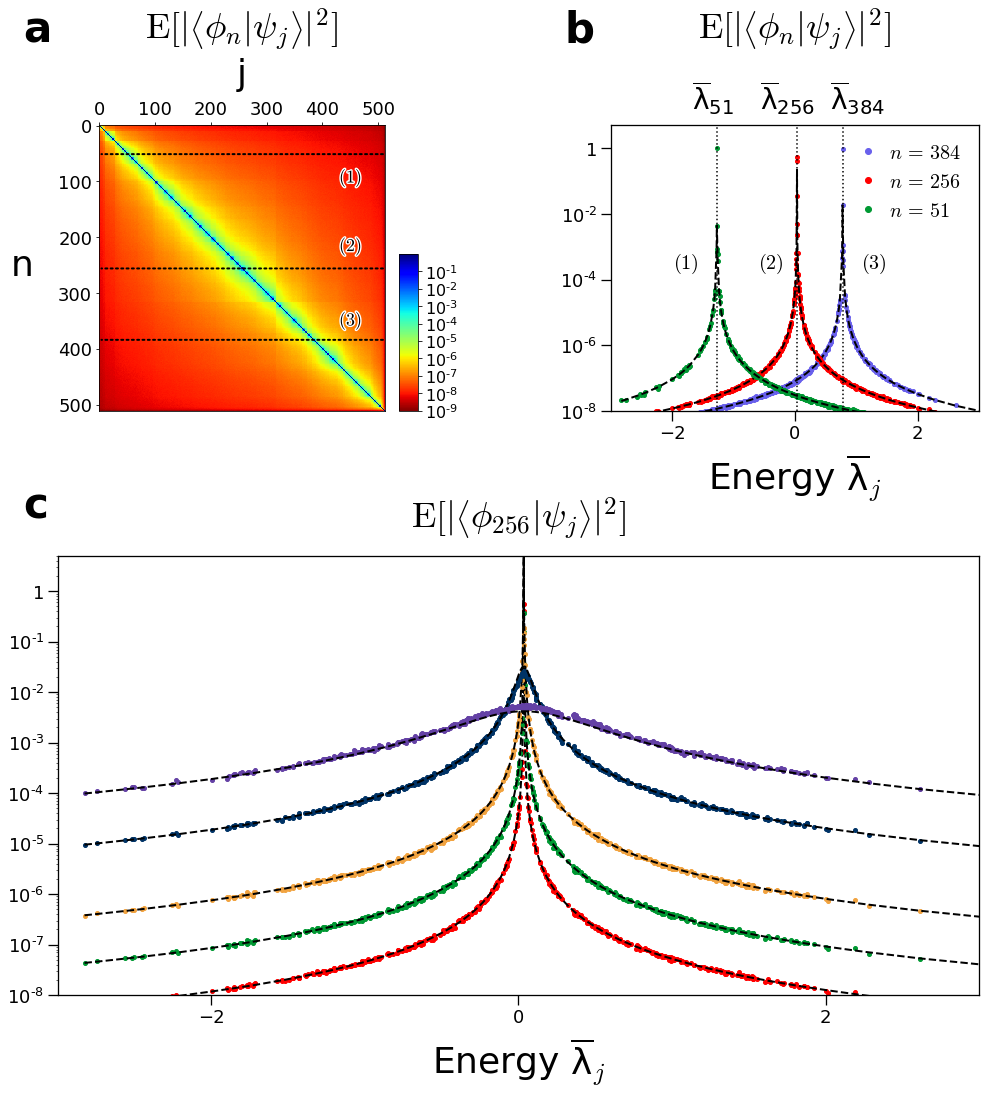}
\caption{ \textbf{Second order moment of the overlaps $\E[| \lan \phi_n | \psi_j \ran |^2]$.}
Numerical simulations are performed with $512\times 512$ matrices, where $\hat{H}_0$ is diagonal with Gaussian distributed eigenvalues of standard deviation $\sigma_0=1$ and zero mean. The interaction Hamiltonian $\hat{W}$ is taken from the GOE ensemble 
with spectral variance  $\sigma_w^2= \Tr(W^2)/N $ ($N=512$).
\textbf{a)} Color plot of the matrix $C_{n,j}=\E[| \langle \phi_n |\psi_j \rangle |^2]$ in the case $\sigma_w=0.01$. This overlap matrix quantifies how much each eigenvector of the ``bare'' Hamiltonian $\hat{H}_0$ is delocalized in the eigenbasis of the dressed Hamiltonian $\hat{H}_0 + \hat{W}$.
The eigenvectors $|\phi_n \rangle$ and $|\psi_i\rangle$ are sorted by order of decreasing eigenvalues.
\textbf{b)}  $\E[| \langle \phi_{n} |\psi_{j} \rangle |^2]$ is plotted as a function of the mean dressed eigenvalue $\bar{\lambda}_j$ of the eigenvector $| \phi_n\rangle$,  for an interaction strength $\sigma_w=0.08$ and several values of $n=51,256,384$ (i.e. one quarter, one half and three quarter of the spectrum respectively).
The theoretical prediction is provided by Eq.\eqref{OverlapSecond} and plotted in dashed line.
\textbf{c)} $\E[| \langle \phi_{256} |\psi_{j} \rangle |^2]$ is plotted as a function of $\bar{\lambda_j}$  and for several values of the coupling $\sigma_w=0.005,0.0135,0.04,0.2,0.65$ (from red to purple). For $\sigma_w=0.005,0.0135$ the regime is clearly perturbative and one has $\E[|\lan \phi_{256}| \psi_{256}|^2] \approx 1$. For $\sigma_w=0.2$ the regime is intermediate and $\Gamma/D = N \pi^2 \sigma_w^2 \rho^2  \approx 31  $, meaning that the eigenvector $|\psi_{256} \ran$ is delocalized over roughly $\approx 30$ bare eigenvectors.
For $\sigma_w=0.65$, the bare eigenvector $|\phi_{256}\ran$ is strongly delocalized.
The theoretical prediction is provided by Eq.(\ref{OverlapSecond}) and plotted in dashed line.
 }
\label{Fig1}
\end{figure}

Combining Eq.\eqref{LinkOverlapResolvent1} and Eq.\eqref{SubordinationResolvent}, we 
get the second order statistics of the overlaps in the case of a Gaussian $\hat{W}$ (either GOE or GUE):
\be
\label{OverlapSecond}
\E[ \lan \phi_n | \psi_i\ran \lan \psi_i | \phi_m\ran ] 
\approx  \f{\delta_{n,m}}{  \rho(\lambda_i)}  l_{\epsilon_n}(\lambda_i)
\quad \text{ with } \quad l_{\epsilon}(\lambda) = \f{1}{\pi} \f{s_\lambda}{(\epsilon-\lambda-\tilde{s}_\lambda)^2+s_\lambda^2 },
\ee
and where 
 $ \tilde{s}_\lambda = \pi \sigma_w^2 H_\rho(\lambda)$ is an energy shift proportional to $H_\rho(\lambda)$ the Hilbert transform of the probability density of the spectrum (i.e. $N \rho$ is the dressed Density  Of States DOS) 
  and
 $s_\lambda = \pi  \sigma_w^2 \rho(\lambda)$ is a decay rate.
 The Lorentzian shape of this formula is reminiscent from the Breit-Wigner law obtained in the context of the so-called ``standard model'' in nuclear physics\cite{BorhMottelson,frazier_strength_1996} and has proved to be ubiquitous in many other fields: 
in molecular physics with the pre-dissociation of diatomic molecules and its effect on rotational absorption lines\cite{brown_predissociation_1932,rice_predissociation_1933}, 
 atomic physics e.g. with the eigenstates properties of the Ce atom\cite{FlaumbaumStructureCompoundState1994},
   thermalisation \cite{deutsch_quantum_1991,genway_dynamics_2013}, 
   financial data analysis with empirical estimation of covariance matrices\cite{LedoitPeche,allez_eigenvectors_2014,bun_rotational_2015,bun_overlaps_2016},
   pure mathematics with free probability\cite{BianeFreeProba,biane_processes_1998,KarginSubordination}.
 It is important to note that, despite the regime is non perturbative, the width $\Gamma=s_\lambda/\pi$ still have a Fermi Golden rule form.
 In addition, in the large energy difference limit 
  the Lorentzian shape reproduces the first order perturbative prediction:
$ \E[|\lan \phi_n | \psi_i \ran|^2] \approx \sigma_w^2/(\epsilon_n -\lambda_i)^2$.
In some sense, Eq.\eqref{OverlapSecond} extends the well known perturbative results valid for $|\epsilon_n -\lambda_i| /\sigma_w \gg 1$ to any value of this ratio.
The predictions obtained are tested numerically on Fig.\ref{Fig1}.

\subsubsection{Generalizations to other statistics on $\hat{W}$}
\label{Generalizations}
 There is a well known generalization of the Gaussian results to Randomly Rotated Matrices 
 i.e. of the form $\hat{W}=\hat{P}.\hat{D}.\hat{P}^\dagger$ with $\hat{D}$ diagonal real fixed and 
$\hat{P}$ unitary or orthogonal Haar distributed. In this case $\hat{H}_0$ and $\hat{W}$ are said to be in generic position with one another, in the sense that the
eigenvectors of $\hat{W}$ are distributed isotropically in the bare eigenbasis. In the limit of infinite dimension, this case provides the framework of \textit{free} probability theory\cite{Voi1,Voi2,BianeFreeProba,biane_conv_sc,VoiculescuFreeConvol}.
It is possible to calculate the mean resolvent of $\hat{H}$ and establish a property of approximate ``subordination'' between $\E[G_{\hat{H}}(z)]$ and $G_{\hat{H}_0}(z)$\cite{KarginSubordination,karginconcentration}:
\be
\label{SubordinationResolvent}
\E[\hat{G}_{\hat{H}_0+\hat{W}}(z)] \approx \hat{G}_{\hat{H}_0}(z+S(z))=\f{1}{\hat{H}_0-z-S(z)}
\ee
where $S(z)$, called the first order subordinate function\cite{SecondOrderSNote}.
 $S(z)$ is related to the analog of a cumulant expansion in a free probability context: the R-transform of
 $\hat{W}$\cite{Mingo1,Mingo2}, $R_{\hat{W}}(z)=\sum_{n \geq 1} \tilde{\kappa}_n(\hat{W}) z^{n-1}$ ($\tilde{\kappa}_n(\hat{W})$ being the free-cumulants) by
 $S(z)= R_{\hat{H}} (m_{\hat{H}}(z))$. 
 The classical cumulants $\kappa_n$ have the property that $\kappa_n(X+Y) = \kappa_n(X)+\kappa_n(Y)$ for two commutative independent random variables $X,Y$. On the free probability side, the free cumulants $\tilde{\kappa}_n$ are such that
 $\tilde{\kappa}_n(\hat{H}_0+\hat{W}) = \tilde{\kappa}_n(\hat{H}_0) + \tilde{\kappa}_n(\hat{W})$ for two \textit{non} commutative random ``free'' variables.
 In some sense, freeness is the equivalent of independence for non commutative random variables.
The Gaussian results from Eq.\eqref{GaussianResolvent} correspond to a truncation of the 
 R-transform at second order
$S(z)=\sigma_w^2 m_{H}(z)$. 
We note in passing the striking resemblance of these results obtained in the framework of free probability theory with the one obtained 
with Dynamical Mean Field Theory (DMFT) in condensed matter\cite{georges_dynamical_1996}. However, it is important to note that DMFT deals with infinite spatial dimension.

\section{Second order statistics of the resolvent and fourth order moments of the overlaps.}
\label{Section4}

In this section, we apply the decoupling technique to the calculation of the fourth order moments of the overlaps for a GUE interaction,
which provides the main result of this paper.
We compare our analytical formulas with numerical simulations on Fig.\ref{Fig2}
and find a satisfactory agreement. 

Again, we use the fact that the overlap coefficient $\lan \phi_n | \psi_i \ran \lan \psi_i | \phi_m \ran$ is (up to a factor $2\pi$) the residue
 of the meromorphic function $G_{n,m}(z) = \sum_j \f{\lan \phi_n | \psi_j \ran \lan \psi_j | \phi_m \ran }{ \lambda_j -z}$ at the pole
  $\lambda_i$. However here, the second order statistics of the resolvent is more complicated, as we will see in the following, after averaging, the function
$(z_1,z_2) \mapsto G_{n,m}(z_1) G_{m,n}(z_2)$ has branch cuts 
and a continuum of singularities (for $z_1=z_2$ in the support of the dressed spectrum).
With a similar procedure as previously done for the first moment of $G_H(z)$, the fourth order moments of the overlap
coefficients can be related to the covariance between Green functions by
%
\be
\label{FourthOrder}
\E[ \lan \phi_n | \psi_i \ran   \lan \psi_i | \phi_m\ran  \lan \phi_p | \psi_j \ran \lan \psi_j | \phi_q \ran]
=
\lim_{\eta,\eta' \to 0^+}
 -\f{\E[\left( G_{n,m}(\bar{\lambda}_i+i\eta)-G_{n,m}(\bar{\lambda}_i-i\eta) \right) 
\left( G_{p,q}(\bar{\lambda}_j +i\eta') -G_{p,q}(\bar{\lambda}_j - i \eta') \right) ]}{4\pi^2  \rho(\bar{\lambda}_i) \rho(\bar{\lambda}_j)} 
\ee
for $i \neq j$.
We are lead to calculate the covariance between Green functions:  $ \E[ G_{n,m}(z_1) G_{p,q}(z_2) ]$.


\subsection{Covariance between Green functions}
%
The covariance between Green functions is also called "diffusion propagator" in the context of the Anderson localization\cite{evers_anderson_2008}.
We apply the same method as in Sec.\ref{LargeZexp} (for spotting the zero cases) and Sec.\ref{FirstStatResolvent} (for calculating the non zero ones).
\subsubsection{Zero covariance cases}
\label{ZCasesOrder2}
A large $|z_1|,|z_2|$ series expansion of both matrix elements $G_{n,m}(z_1)$ and $G_{p,q}(z_2)$ is made, and then the Wick theorem is used (the Weingarten calculus can be used for orthogonal or unitary Haar distributed interactions, see Supp. Mat. of \cite{ithier_typical_2017}).
 $\E[G_{n,m}(z_1) G_{p,q}(z_2)]$ is zero $\forall z_1,z_2$, 
except if
\begin{itemize} 
\item  ($n=m$ and $p=q$): these are correlations between the diagonal terms of the resolvent, $\E[G_{n,n}(z_1)G_{p,p}(z_2)]$.
These correlations are involved in the time evolution of the coherence terms of the total density matrix $\varrho(t)$ 
 under the Hamiltonian $\hat{H}$ (i.e. the \textit{extra} diagonal terms of $\varrho(t)$).
\item ($n=q$ and $m=p$): correlations between the two extra diagonal entries $\E[G_{n,m}(z_1)G_{m,n}(z_2)]$.
These correlations are involved in the time evolution of the \textit{diagonal} terms of $\varrho(t)$.
\end{itemize}
To calculate the non zero cases: $\E[G_{n,m}(z_1) G_{m,n}(z_2)]$ and $\E[G_{n,n}(z_1) G_{m,m}(z_2)]$, we proceed using the loop equation method.
\subsubsection{Calculation method for the non zero cases}
\label{CovGreenFunc}
In the following, we will focus on a unitary Gaussian interaction $W$.
We apply the same method as for the first order statistics case: 
\begin{itemize}
\item expand  both Green functions on the right and on the left using Eq.\eqref{ResolventEntryExpansion},
\item consider the average over the statistics of $W$ and use the decoupling formula from Eq.\eqref{WG},
\item assume that the Stieltjes transform of the spectrum of $\hat{H}$ is concentrated and neglect the fluctuations 
(one can justify this approximation using Corollary 4.4.30 in \cite{agz}),
\item use the relation $G_H(z_1) G_H(z_2)= -(G_H(z_1)-G_H(z_2))/(z_1-z_2)$.
\item finally get a set of self-consistent approximate equations verified by the covariance $\E[G_{n,m}(z_1) G_{p,q}(z_2)] $,
\end{itemize}
This procedure, described in detail in the Appendix, provides a set of four 
Schwinger-Dyson equations (or loop equations, see \cite{GuionnetLargeRandomMat} and also Chap. 6 in \cite{EynardLecture})
which can be combined together to get 
the following results. 
%
\subsubsection{Covariance betwen Green functions}
\begin{itemize}
\item The covariance between extra diagonal Green functions is
\begin{eqnarray}
\label{CovExtraDiag}
\boxed{\E[G_{n,m}(z_1) G_{m,n}(z_2)]
\approx   \frac{\sigma_w^2}{N}  \left(1+ S_2(z_1,z_2) \right) F_{n,m}(z_1,z_2)  }\\
 \\
 \text{with } \begin{cases} F_{n,m}(z_1,z_2)= \E[G_{m,m}(z_1) ] \E[G_{m,m}(z_2)] \E[G_{n,n}(z_1)] \E[G_{n,n}(z_2)]  \\ \\
   S_2(z_1,z_2)=\sigma_w^2 \dfrac{m_H(z_1)-m_H(z_2)}{z_1-z_2} = \sigma_w^2 \tr(\E[G_{H}(z_1) G_{H}(z_2)]).  
 \end{cases}
\end{eqnarray}
These equations provide a direct link between the first and second order statistics of the resolvent. By analogy to the subordinate function involved in the first 
order statistics of the resolvent of $H+W$ (i.e. $S(z)=\sigma_w^2  \tr(\E[G_{H}(z)])$ when $W$ is unitary Gaussian),
 the function $S_2(z_1,z_2)$ involved in Eq.\eqref{CovExtraDiag} can be named a ``second order'' subordinate function.
This function has a continuum of singularities for $z_1 =z_2$ in the dressed spectrum.

\item The covariance between diagonal entries is given by
\begin{eqnarray}
\label{CovDiag}
\boxed{
  \E[\delta G_{n,n}(z_1) \delta G_{p,p}(z_2)]  
\approx -   \frac{\sigma_w^2}{N} \f{1}{z_2-z_1}
\f{\E[(G_{p,n}(z_2)-G_{p,n}(z_1)) (G_{n,p}(z_2)-G_{n,p}(z_1))] }{ \epsilon_n-\epsilon_p +z_2 -z_1 +\sigma_w^2 (m_H(z_2)-m_H(z_1))
}
}.
 \end{eqnarray}
 \end{itemize}
 meaning that it smaller by a factor 
  $\sigma_w^2/N$ compared to 
 the extra diagonal covariance.

\subsection{Fourth order statistics of the overlaps}
\label{FourthOrderStatOverlap}
 %

%
\subsubsection{Case $n=q$, $m=p$: $\; \E[ \lan \phi_n | \psi_i \ran   \lan \psi_i | \phi_m\ran
\lan \phi_m | \psi_j\ran \lan \psi_j | \phi_n\ran ]$}
\begin{figure}
\includegraphics[width=0.8\textwidth]{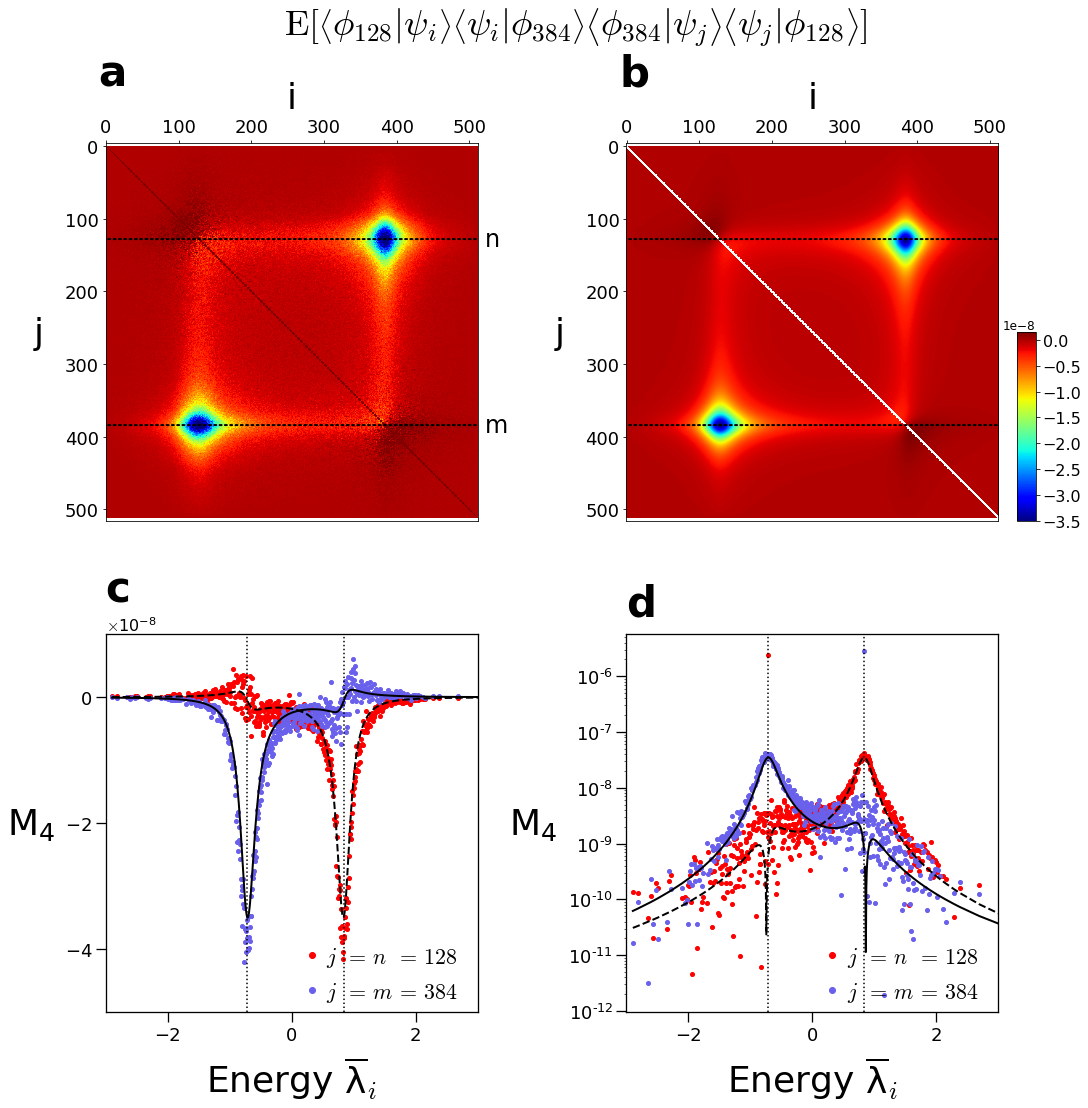}
\caption{ \textbf{Fourth order moments of the overlaps: $C_{i,j}=\E[ \langle \phi_n | \psi_i \rangle.\langle \psi_i | \phi_m \rangle
.\langle \phi_m  |\psi_j \rangle. \langle \psi_j | \phi_n \rangle ]$.}
Here $\hat{H}_0$ is a $N\times N$ ($N=512$) diagonal matrix with a centered gaussian DOS with variance $1$, $\hat{W}$ is in the GOE with 
 $\sigma_w=\Tr(W^2)/N=0.4$. 
 \textbf{a)} Color plot of the matrix $C_{i,j}$ for fixed values of $n=128=N/4$ and $m=384=3N/4$. 
\textbf{b)}  Theoretical predictions from Eq.(\ref{OverlapDiagTheo}).
  After Fourier transform, these overlaps give the diagonal part of average total density matrix $\E[\rho(t)]$, i.e they provide the out of equilibrium dynamics of the probabilities of occupation of a quantum system.
\textbf{c)} $C_{i,j}$ is plotted as a function of $\bar{\lambda}_i$ for $j=n=128$ (red) and for $j=m=384$ (blue) with linear scale.
\textbf{d)} Same plot in log scale (of $|C_{i,j}|$). 
The averaging is performed over $8.10^5$ realizations of $\hat{W}$.
%
}%
\label{Fig2}
\end{figure}

Combining Eq.\ref{FourthOrder} and Eq.\ref{CovExtraDiag}, we get the main result of this article, the fourth order moment of the overlaps for $i \neq j$ and $n\neq m$:
\begin{equation}
\boxed{
 \E[ \lan \phi_n | \psi_i \ran   \lan \psi_i | \phi_m\ran
\lan \phi_m | \psi_j\ran \lan \psi_j | \phi_n \ran ] \approx 
- \f{\sigma_w^2}{N}  \f{1}{ \rho(\lambda_i)\rho(\lambda_j)} 
\f{ l_{\epsilon_n}(\lambda_i)l_{\epsilon_m}(\lambda_i) - l_{\epsilon_n}(\lambda_j) l_{\epsilon_m}(\lambda_i)  }{(\lambda_i -\lambda_j)(\epsilon_n- \epsilon_m)}}
\label{OverlapDiagTheo}
\end{equation}
where $l_{\epsilon} (\lambda)$ is the Lorentzian function we introduced in Eq.\eqref{OverlapSecond}.  We have assumed $\rho(\lambda_i,\lambda_j) \approx \rho(\lambda_i) \rho(\lambda_j)$ when $i\neq j$. This formula is tested numerically on Fig.\ref{Fig2} and we find a satisfactory agreement.
We will use this formula when studying the out of equilibrium dynamics of an embedded quantum system\cite{IthierTransient}.

\subsubsection{Case $n=m$ and $p=q$: $\; \E[| \lan \phi_n | \psi_i \ran|^2    | \lan \psi_j | \phi_p\ran|^2 ]$}
As we saw in Sec., two distinct diagonal entries of the resolvent are weakly correlated. As a result the
fourth order moment of the overlap follows easily:
\begin{eqnarray*}
\E[\left( G_{n,n}(z_1) -G_{n,n}(z_1^*)\right) \left( G_{p,p}(z_2) -G_{p,p}(z_2^*) \right) ]
& = &-4 \E[\Im(G_{n,n}(z_1)) \Im(G_{p,p}(z_2))] \approx -4 \Im(\E[G_{n,n}(z_1)]) \Im(\E[G_{p,p}(z_2)])
\end{eqnarray*}
Using Eq.\ref{FourthOrder}
we get the overlap moment:
\be
\label{Overlap2A}
\boxed{
\E[ |\lan \phi_n | \psi_i \ran |^2  |\lan \phi_p | \psi_j\ran |^2 ]
\approx \E[ |\lan \phi_n | \psi_i \ran |^2 ] \E[ |\lan \phi_p | \psi_j\ran |^2 ]
\approx \f{1}{ \rho(\lambda_i)\rho(\lambda_j)} l_{\epsilon_n}(\lambda_i) l_{\epsilon_p}(\lambda_j)
}
\ee
with the Lorentzian functions $l_\epsilon(\lambda)$ defined in Eq.\ref{OverlapSecond}.

This formula works for all cases, except for the case ($n=p$ and $i=j$).
 
 \section{Conclusion}
 We presented a method for calculating the mixed moments of overlaps between eigenvectors of two large matrices: $\hat{H}_0$ deterministic arbitrarily
chosen and $\hat{W}$ random. We applied this method to calculate the second and fourth order moments in the Gaussian case for $\hat{W}$.
These quantities are crucial for understanding the out of equilibrium dynamics of an embedded quantum system.
The formulas we obtain were tested numerically and a satisfactory agreement was found.
The method presented in the article will be used for investigating other statistics for $\hat{W}$: Wigner Random Band Matrices, matrices with correlated entries, and Embedded ensembles in a further publication. 


 Acknowledgements: we would like to thank F. Benaych-Georges for insightful discussions and
 spotting to us useful references.

\include{SupMatOverlaps}
\bibliography{Thermalisation}

\begin{thebibliography}{85}
\bibitem [{Fre()}]{FreeProbaNote}%
  \BibitemOpen
  \href@noop {} {}\bibinfo {note} {By generic position, we mean that the
  eigenvectors of one matrix are distributed isotropically in the eigenbasis of
  the other.}\BibitemShut {Stop}%
\bibitem [{Gre()}]{GreenFunctionNote}%
  \BibitemOpen
  \href@noop {} {}\bibinfo {note} {We call Green functions the matrix elements
  of the dressed resolvent operator in the bare basis.}\BibitemShut {Stop}%
\bibitem [{Sec()}]{SecondOrderSNote}%
  \BibitemOpen
  \href@noop {} {}\bibinfo {note} {A second order subordinate function will be
  considered when dealing with the covariance between Green functions in
  Sec.\ref{Section4}}\BibitemShut {NoStop}%
\end{thebibliography}

\end{document}